\documentclass[conference]{IEEEtran}
\IEEEoverridecommandlockouts

\usepackage{orcidlink}
\usepackage{cite}
\usepackage{caption}
\usepackage{amsmath,amssymb,amsfonts}
\usepackage{algorithmic}
\usepackage{graphicx}
\usepackage{textcomp}
\usepackage{xcolor}
\usepackage{comment}
\usepackage{array}
\usepackage{tabularx}
\usepackage{booktabs} 
\usepackage{listings}
\usepackage{caption}
\usepackage{mathtools} 
\usepackage[T1]{fontenc}
\usepackage{float} 
\usepackage{units}
\def\BibTeX{{\rm B\kern-.05em{\sc i\kern-.025em b}\kern-.08em
    T\kern-.1667em\lower.7ex\hbox{E}\kern-.125emX}}
\usepackage{hyperref}
\hypersetup{
pdftitle={Verifiable Decentralized IPFS Cluster: Unlocking
Trustworthy Data Permanency for Off-Chain
Storage},
pdfauthor={Sid Lamichhane, Patrick Herbke},
pdfkeywords={Decentralized Applications, Blockchain, Off-
chain storage, InterPlanetary File System (IPFS), IPFS Cluster,
Decentralized Identifiers (DIDs), Verifiable Credentials (VCs),
Data Permanency, Trust, Privacy, Security}
}

\begin{document}

\title{Verifiable Decentralized IPFS Cluster:\\ Unlocking Trustworthy Data Permanency for Off-Chain Storage}

\author{\IEEEauthorblockN{
Sid Lamichhane \orcidlink{0009-0000-5845-7060}\IEEEauthorrefmark{1},
Patrick Herbke \orcidlink{0000-0001-9649-2975}\IEEEauthorrefmark{1}
}
\IEEEauthorblockA{\IEEEauthorrefmark{1}Service-centric Networking,
Technische Universit\"at Berlin, Berlin, Germany\\
\{lamichhane, p.herbke\}@tu-berlin.de}}

\maketitle

\begin{abstract}
In Decentralized Applications, off-chain storage solutions such as the InterPlanetary File System (IPFS) are crucial in overcoming Blockchain storage limitations. However, the assurance of data permanency in IPFS relies on the pinning of data, which comes with trust issues and potential single points of failure. This paper introduces Verifiable Decentralized IPFS Clusters (VDICs) to enhance off-chain storage reliability with verifiable data permanency guarantees. VDICs leverage Decentralized Identifier, Verifiable Credentials, and IPFS Clusters to create a trustworthy ecosystem where the storage of pinned data is transparent and verifiable. Performance evaluations demonstrate that VDICs are competitive with traditional pinning services. Real-life use cases validate their feasibility and practicality for providers of Decentralized Applications focused on ensuring data permanency.
\end{abstract}

\begin{IEEEkeywords}
Decentralized Applications, Blockchain, Off-chain storage, InterPlanetary File System (IPFS), IPFS Cluster, Decentralized Identifiers (DIDs), Verifiable Credentials (VCs), Data Permanency, Trust, Privacy, Security
\end{IEEEkeywords}

\section{Introduction}
Decentralized Applications (DApps) are Blockchain-based software systems adopted across diverse domains such as finance, gaming, and health care~\cite{10068327}. DApps commonly utilize off-chain storage to overcome limitations in Blockchain storage capacity. The InterPlanetary File System (IPFS) is often chosen for off-chain storage due to its immutability and data permanency~\cite{9903549, 9027313, JAYABALAN2022152}. 

However, data permanency in IPFS is guaranteed only when data is pinned. Pinning is an inherent functionality of IPFS that labels data as pinned to prevent its erasure by the automatic deletion process in IPFS. Pinning can be achieved either by running a personal IPFS node or by using pinning services offered by third parties for financial compensation~\cite{ipfsPersistenceIPFS}. Both options have distinct drawbacks: running a personal node introduces a single point of failure, while using pinning services necessitates trust~\cite{9369473, 9903549}. Trusting pinning service providers is explicitly cautioned against by the open-source IPFS community~\cite{ipfsPersistenceIPFS}.

While pinning guarantees data permanency in IPFS from a technical standpoint, the transparency and verifiability of the entities storing the pinned data in IPFS remain issues. IPFS is maintained by anonymous node operators whose motives for permanently storing data in their local infrastructure are not identifiable or verifiable by anyone~\cite{9369473}. This lack of insight into IPFS node operators raises questions about why they would bear the expense of permanently storing others' data without apparent incentives. Consequently, the guarantee of data permanency by IPFS is questionable.

\textbf{Contribution. }
With Verifiable Decentralized IPFS Clusters (VDICs), we present a novel solution for storing DApp data off-chain. VDICs fulfill the same function as IPFS and offer an enhancement by providing verifiable guarantees of data permanency. The verifiability of data permanency empowers users and providers of DApps to confidently assess data storage's reliability in VDICs. Consequently, VDICs have the potential to revolutionize off-chain storage, ensuring trustworthy data permanency for any DApp.
To evaluate VDICs, we conducted performance tests against traditional pinning services, demonstrating the competitiveness of our solution. Furthermore, real-life use cases validate the feasibility and practicality of VDICs, strengthening their potential impact in DApps.

\section{Related Work}
This section reviews the current state of decentralized storage solutions that guarantee data permanency.

Filecoin~\cite{filecoinWebsite} is a decentralized storage network that extends IPFS by providing economic incentives and cryptographic proofs to provide verifiable permanent off-chain storage for DApps. Although VDICs also provide verifiable permanent off-chain storage, our approach differs. 
One difference is the method of guaranteeing data permanency. While Filecoin relies on financial incentives, VDICs differ by not requiring DApps to pay for data permanency, enhancing DApp reliability by avoiding failures due to exhausted funds.
Another difference lies in the verification of data permanency. Filecoin employs self-created proof systems, whereas VDICs utilize Verifiable Credentials (VCs) for verifying data permanency~\cite{Benet2017PoRep, proofOfSpacetime}. VCs are standardized, digitally signed claims that can be independently verified. The ongoing research and numerous tested use cases of VCs enhance the reliability of our verification method, fostering trust in our solution~\cite{8869262, 9031543}. 
Furthermore, Filecoin faces limitations that VDICs address effectively. One such limitation is the complexity of creating storage space~\cite{ipfsPersistenceIPFS}. VDICs simplify storage space creation by offering built-in APIs for DApps, facilitating a less complex integration and workflow for DApp providers. Additionally, VDICs provide data privacy and authentication-based data access, whereas all data in Filecoin is publicly accessible. Offering private and public data availability makes VDICs more adaptable to the varying needs of DApps.

Muralidharan et al.~\cite{8662002} explore the potential of decentralizing the management of Internet of Things data using IPFS Clusters, demonstrating feasibility and advantages through experimental evaluations on Raspberry Pi devices. Muralidharan et al. utilize IPFS Clusters to create decentralized storage networks but miss the verifiability of its IPFS Cluster node operators. Verifying node operators in VDICs allows the public to assess their guarantee of data permanency, leading to trustworthy data permanency. Therefore, our solution enhances the decentralized storage network concept introduced by Muralidharan et al. and improves it by incorporating verifiable guarantees of data permanency.

Steichen et al.~\cite{8726493} propose a modified version of IPFS that integrates Ethereum smart contracts to enable access-controlled file sharing while ensuring privacy and security. Their approach uses smart contracts to maintain an access control list, regulating file access based on permissions recorded on the Ethereum Blockchain. 
However, their solution is limited by the public nature of IPFS, eventually exposing data once it is replicated by IPFS nodes outside their proposed file-sharing system. In contrast, we use private IPFS Clusters to ensure that data remains private within a controlled environment, reducing the risk of unauthorized access and replication. 
Furthermore, our solution utilizes VCs instead of smart contracts for access control. By using VCs, we leverage from existing authentication protocols, such as OpenID for Verifiable Credentials~\cite{oid4vc} or OpenID Connect~\cite{openidOpenIDConnect}, eliminating the need for implementing new access authentication schemata based on smart contracts. Moreover, using well-established authentication standards simplifies the integration of VDICs and enhances their security.
By combining robust access control mechanisms with private IPFS Clusters, VDICs offer enhanced, privacy-focused decentralized data storage, improving upon the approach proposed by Steichen and colleagues.

Despite the advancements in existing research, there remains a critical gap in providing off-chain storage that combines data permanency, verifiability, and privacy. Our work addresses these challenges by offering VDICs as an alternative that ensures trustworthy data permanency in private off-chain storage of DApps.

\section{Background} 
Our solution leverages IPFS Cluster, Decentralized Identifiers, and Verifiable Credentials to provide a framework for verifiable permanent off-chain storage, integrable with any DApp. By using Linked Verifiable Presentations, VDICs and their guarantee of data permanency are verifiable by the public, bringing trust off-chain.

\subsection{InterPlanetary File System and IPFS Cluster}
The InterPlanetary File System (IPFS) is a decentralized protocol designed to create a permanent and distributed method for storing and sharing content on the internet. IPFS employs a content-addressable system, where files are identified by unique cryptographic hashes rather than their location on a server. Content-addressing allows for efficient content distribution because files can be retrieved from any node in the network rather than relying on centralized servers~\cite{ipfstech}. 

IPFS Cluster extends the capabilities of IPFS by providing a mechanism for orchestrating and managing multiple IPFS nodes in a clustered environment. Clustering IPFS nodes enables scalability, reliability, and fault tolerance for storing and accessing data on the IPFS network. Therefore, IPFS Cluster suits applications requiring high availability and data redundancy. In particular, IPFS Clusters power large IPFS storage services such as the pinning service of web3.storage~\cite{ipfscluster}.

\subsection{Decentralized Identifiers}
Decentralized Identifiers (DID) are a fundamental component of decentralized identity systems, providing a mechanism to uniquely identify entities in a decentralized and interoperable manner~\cite{w3c-did-core}. DIDs are globally unique identifiers that are cryptographically secure and independent of any centralized registry or authority. DIDs enable entities, such as individuals, organizations, or things, to assert control over their digital identity by providing a persistent, self-sovereign means of identification. DIDs typically consist of a method-specific identifier string, which is combined with a method-specific DID scheme, allowing flexibility and extensibility across various decentralized identity systems. By leveraging DIDs, entities can establish trust relationships, authenticate themselves, and interact with DApps while maintaining privacy and control over their personal information.

\subsection{Verifiable Credentials}
Verifiable Credentials (VCs) are digital representations of attestations, assertions, or qualifications about a subject, such as an individual, organization, or thing~\cite{w3c-vc-data-model}. These credentials are cryptographically secured and can be shared between entities over the internet. They enable verifiable proof of claims without the need for a central authority, fostering decentralized trust and interoperability in digital interactions. VCs adhere to principles of privacy and user control, empowering individuals to manage their own digital identity and selectively disclose information as needed.

\subsection{Linked Verifiable Presentation}
Linked Verifiable Presentations, introduced by the Decentralized Identity Foundation, are a method for publicly sharing, discovering, and retrieving VCs using service entries in DID Documents. To publicly share VCs, Linked Verifiable Presentations require holders to provide Uniform Resource Locators, pointing to the Verifiable Presentation of their VCs, e.g., stored in IPFS or self-hosted servers. Linked Verifiable Presentations complement private communication protocols such as OpenID for Verifiable Credentials, which are used for the private sharing of VCs~\cite{identityLinkedVerifiable}.

Verification of Linked Verifiable Presentations begins with verifiers examining the digital signatures on each credential using the public key of the credential's issuer, which is obtained from the issuer's DID Document. Through this process, verifiers confirm the authenticity of the credential, confirming its issuance by the claimed issuer without any alterations. Subsequently, verifiers validate the holder's DID by ensuring the authenticity and integrity of the associated DID Document, often through Blockchain or other decentralized ledger technologies~\cite{identityLinkedVerifiable}.

\section{Verifiable Decentralized IPFS Clusters}
Verifiable Decentralized IPFS Clusters (VDICs) provide off-chain storage for DApps. VDICs are IPFS Clusters that are (1) decentralized due to multiple different IPFS Cluster node operators and (2) verifiable by anyone due to the integration of VCs.

In this section, we first describe actors in VDICs. Secondly, we outline the architecture of VDICs. Thirdly, we explain how decentralization and verifiability establish trustworthy data permanency in VDICs. Finally, we present a method for implementing VDICs.

\begin{figure}[!h]
    \centering
    \includegraphics[width=\linewidth]{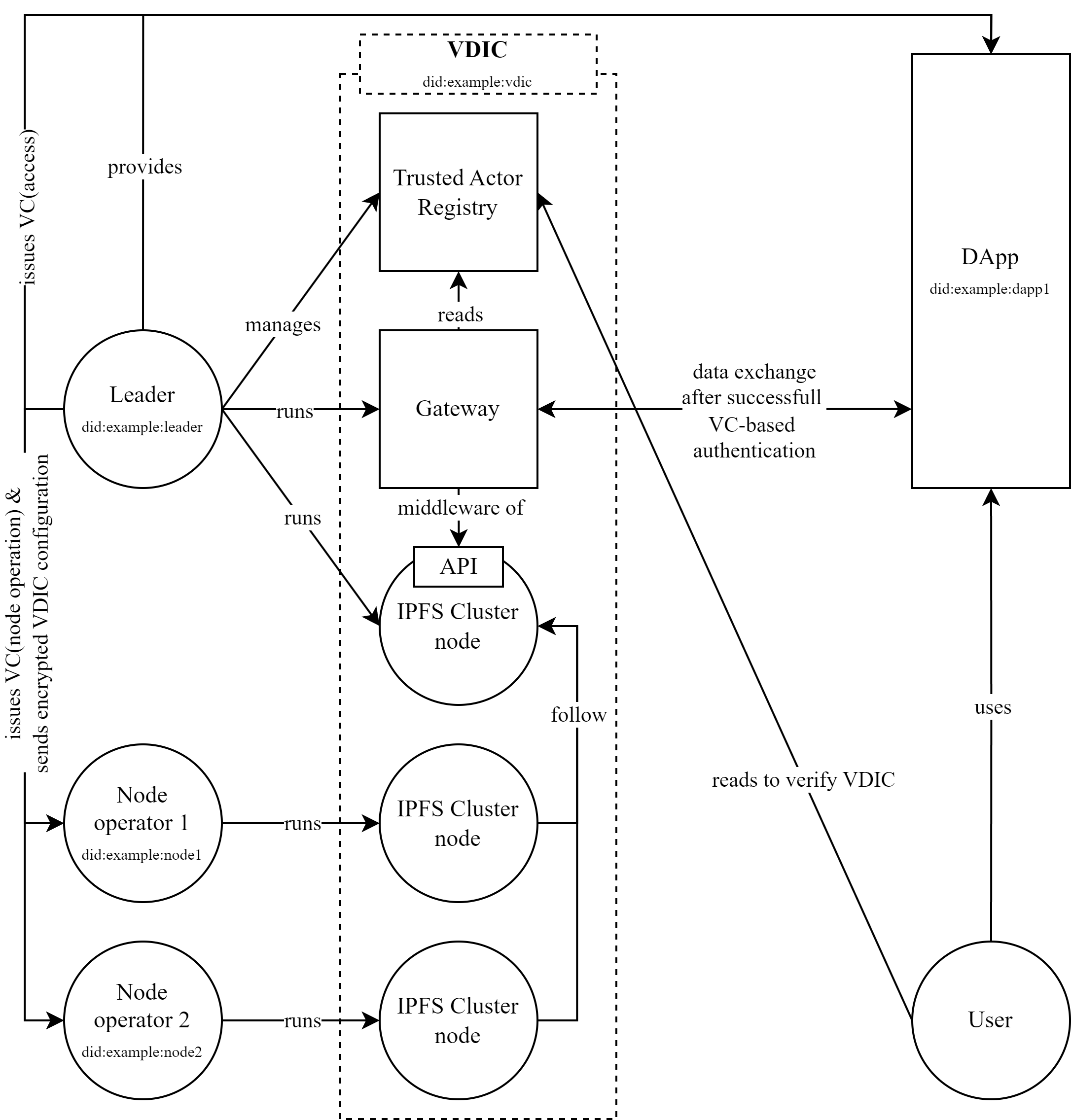}
    \caption{Interplay between actors (leaders, node operators, DApps, and DApp users) and architectural components of VDICs (Trusted Actor Registry, Gateway, and IPFS Cluster)}
    \label{fig:network_architecture}
\end{figure}

Figure~\ref{fig:network_architecture} shows the interplay between the actors and the architecture of VDICs, characterizing VDICs as sociotechnical networks due to the integration of human and technological elements. 

\subsection{Actors}
VDICs connect diverse actors with each other. Actors have clear rights and obligations within VDICs, further detailed in the following:

\begin{itemize}
    \item \textit{Leaders} are single persons, groups, or legal entities. They provide the DApp that uses the VDIC they lead. 
    
    The obligations of leaders are manifold, highlighting their pivotal role. First, leaders run the essential infrastructure of VDICs, such as the Gateway component and at least one IPFS Cluster node. Second, leaders must possess a DID that enables them to issue VCs for authorizing other actors in VDICs. Third, leaders must document the authorized actors in the Trusted Actor Registry component of their VDIC.

    Leaders have the right to authorize anyone to act in their VDIC as node operator or DApp. However, leaders must scrutinize actors in a self-defined application process before giving authorization.
    
    \item \textit{Node operators} are single persons, groups, or legal entities. They run IPFS Cluster nodes in VDICs, giving them the right to read and use all data in VDICs.
    
    Node operators are obligated to possess a DID listed in the Trusted Actor Registry component of their VDIC. The DID of node operators must resolve to one DID Document containing VCs that prove their identity. The DID Document must be publicly available to enable anyone to verify node operators and, thus, VDICs.
    
    \item \textit{DApps} are accessors of VDICs. Their access authorization is issued as VCs by leaders. Depending on their access authorization, DApps have the right to either read and write data in VDICs or only read.
    
    DApps are obligated to possess a DID listed in the Trusted Actor Registry component of their VDIC. The DID of DApps must resolve to one DID Document containing VCs that prove their provider's identity. The DID Document must be publicly available to enable anyone to verify DApps and, thus, VDICs.
    
    \item \textit{DApp users} are verifiers of VDICs. DApp users want to ensure that the data they produce while using DApps connected to VDICs is permanent. To verify VDICs and their guaranteed data permanency, DApp users rely on the Trusted Actor Registry component as explained in section~\ref{subsec:data_permanency}.
\end{itemize}

\subsection{Architecture}
VDICs are IPFS Clusters extended by the components: Trusted Actor Registry and Gateway. This section describes both components and their purposes.

\subsubsection{Trusted Actor Registry} 
Each VDIC possesses a DID, which is controlled by its leader. This DID resolves to one DID Document, which includes the Trusted Actor Registry of the respective VDIC. Trusted Actor Registries list the DID of authorized node operators and DApps. Therefore, reading Trusted Actor Registries is the first step in verifying VDICs and their actors.
 
Verification of VDICs must be possible by anyone; thus, Trusted Actor Registries must be stored in public Blockchains to be publicly available. Therefore, the DID of a VDIC must be based on a DID method that relies on public Blockchains. Such DID methods are, for example, did:cheqd, did:ebsi, or did:ethr.

Listing~\ref{lst:trusted_did_registry} reveals an exemplary implementation of Trusted Actor Registries in the service section of DID Documents, using two service entries of type Linked Domains. While one service entry lists the DID of authorized node operators, the other entry records the DID of DApps with access authorization.

\begin{lstlisting}[label=lst:trusted_did_registry, caption=Trusted Actor Registry implemented inside service section of DID Document]
{
    "@context": ["https://www.w3.org/ns/did/v1"],
    "id": "did:example:vdic_id",
    "controller": ["did:example:leader"],
    "verificationMethod": [...],
    "authentication": [...],
    "service": [
        {
            "id": "did:example:vdic#node_operators",
            "type": "LinkedDomains",
            "service endpoint": [
                did:example:node_operator_1, 
                did:example:node_operator_2, 
                ..., 
                did:example:node_operator_n
            ]
        },
        {
            "id": "did:example:vdic#dapps",
            "type": "LinkedDomains",
            "service endpoint": [
                did:example:dapp_1, 
                did:example:dapp_2, 
                ...,
                did:example:dapp_m
            ]
        }
    ]
}
\end{lstlisting}

\subsubsection{Gateway} 
Gateways enable authorized DApps to read and write data in VDICs. Gateways function as middleware of the default API of the IPFS Cluster node, run by VDIC leaders. As middleware, Gateways add a VC-based authentication system to the default API that allows anyone to read and write data to IPFS Clusters~\cite{ipfsclusterREST}.

Authentication is needed for any DApp prior read or write operations in VDICs. Access is granted after DApps present their VC with access claims to Gateways. The presented VCs are additionally examined by Gateways regarding revocation. To check the revocation status of VCs, Gateways read the Trusted Actor Registry and confirm whether the DID of the access-requesting DApp is listed. If DApps are not listed in the Trusted Actor Registry, their access VC was revoked by VDIC leaders.

\subsection{Trustworthy data permanency}
\label{subsec:data_permanency}
VDICs provide trustworthy permanency of their data that is produced by DApps using VDICs as off-chain storage. Providing trustworthy data permanency requires VDICs (1) to ensure data permanency technically and (2) to create trust in their guarantee for data permanency.

Data permanency is attained through decentralization. By decentralizing IPFS Clusters, VDICs achieve data replication across multiple independent nodes, thereby enhancing fault tolerance and reliability. Therefore, the level of decentralization in VDICs determines their technical ability to guarantee data permanency, i.e., the higher the number of distinct node operators in VDICs, the higher the chance of VDICs providing data permanency.

Trust in the guarantee for data permanency is attained through verifiability. By verifying the number of nodes in VDICs and each node operator's identity and motive for storing data permanently, the public can scrutinize the data permanency guaranteed by VDICs. To scrutinize and thus gain trust in the guaranteed data permanency, one must first retrieve the publicly available Trusted Actor Registry of VDICs. Trusted Actor Registries are Blockchain-based lists that allow only VDIC leaders to record the DID of node operators. Resolving the DID of node operators to its DID Document reveals claims about their identity and motive for storing data permanently. These claims are stored as Linked Verifiable Presentations, ensuring public verifiability~\cite{identityLinkedVerifiable}. To ensure the existence of these Linked Verifiable Presentations, leaders must require all node operators to provide these presentations before authorizing the operation of nodes in VDICs. 

\begin{figure}[!h]
    \centering
    \includegraphics[width=\linewidth]{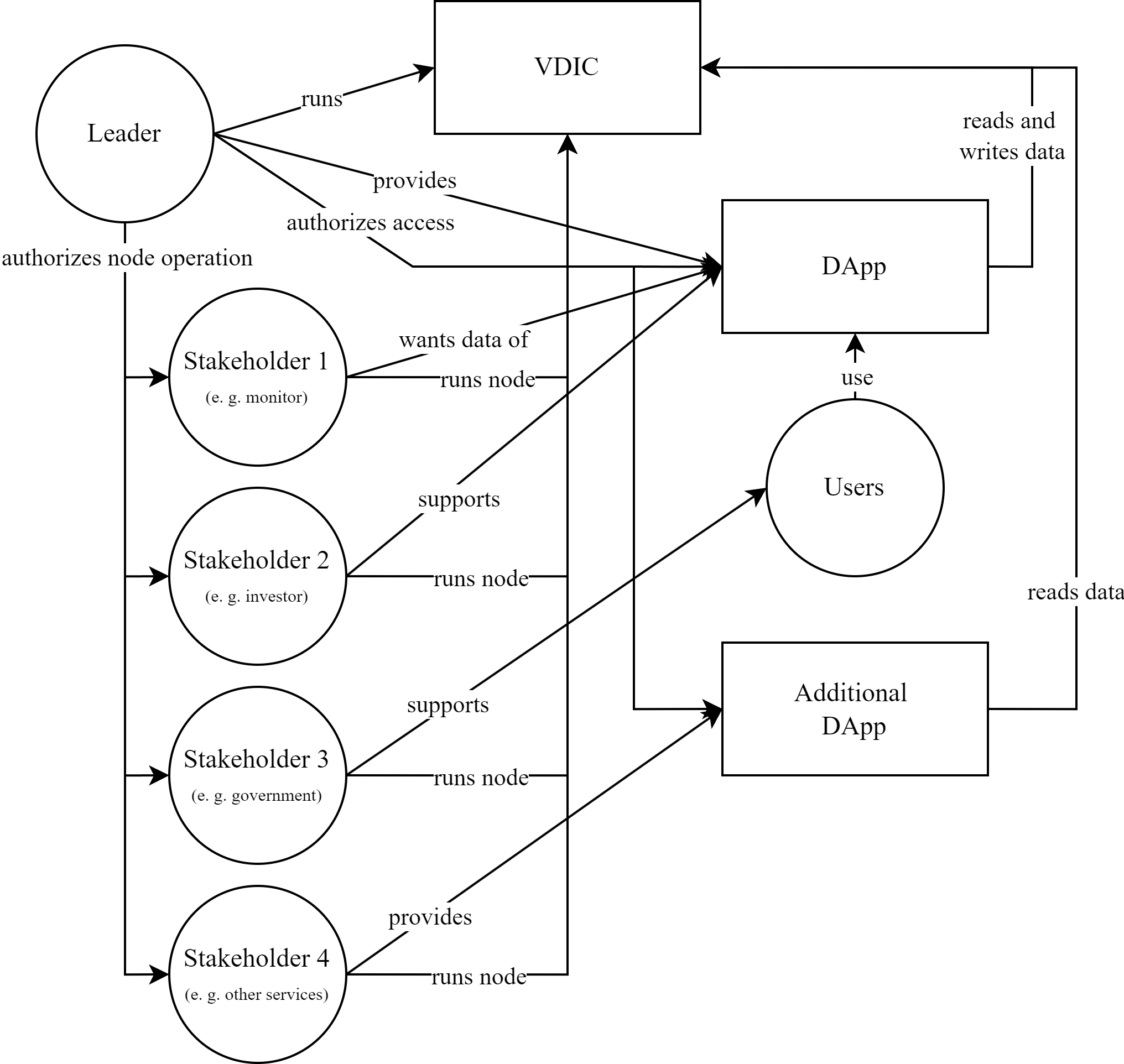}
    \caption{Relationships between actors of VDICs and the four general motives for node operation: data availability (stakeholder 1), support of DApps using VDICs as off-chain storage (stakeholder 2), support of users relying on services provided by DApps using VDICs (stakeholder 3), and provision of DApps using VDICs as data sources (leader and stakeholder 4)}
    \label{fig:stakeholders}
\end{figure}

The motive of node operators to store data permanently may be decisive for the public to trust the data permanency guaranteed by VDICs. Additionally, the presence of motivated node operators supports DApp providers in justifying the use of VDICs for off-chain storage. Therefore, leaders should require node operators to have a motive to join VDICs and keep data permanently. Figure~\ref{fig:stakeholders} summarizes the four general motives of node operators in VDICs:

\begin{itemize}
    \item \textit{Data availability} motivates node operators to keep data permanently. Running one's own node yields the highest possible data availability. Therefore, stakeholders monitoring or analyzing data stored in VDICs in real-time should experience this motivation for data availability.
    
    \item \textit{Support of DApps using VDICs as off-chain storage} motivates node operators to keep data permanently. DApps that use VDICs as off-chain storage may have stakeholders, e.g., investors or partners. Such stakeholders might be willing to run nodes to support further the DApp in which they have a stake.
    
    \item \textit{Support of users relying on the service provided by DApps using VDICs} motivates node operators to keep data permanently. The service provided by DApps using VDICs might benefit users who other institutions such as NGOs, associations, and governments also want to support. Therefore, such institutions might be willing to run nodes and thus ensure the continuity of the provided service.

    \item \textit{Provision of DApps using VDICs as data sources} motivates node operators to keep data permanently. The success of DApps that use VDICs as data sources relies on the data permanency guaranteed by VDICs. Therefore, providers of such DApps are motivated to operate nodes. Moreover, providers might be further motivated because of the high data availability their own nodes can provide for their DApps.
\end{itemize}

\subsection{Implementation}
Our implementation of VDICs is based on private IPFS Clusters, providing immutable, permanent, and private off-chain storage for any DApp. Further configurations of VDICs also allow for public IPFS Clusters. This section outlines how to create, update, use, and configure our implemented VDIC.

\subsubsection{Create VDICs}
To create VDICs, leaders must first generate a DID for their VDIC to initialize an empty Trusted Actor Registry. Second, leaders create an IPFS Cluster using the command line application ipfs-cluster-service~\cite{ipfsclusterIpfsclusterservice}. The created IPFS Cluster is configured by adapting the default IPFS Cluster configuration file. To run the configured IPFS Cluster, leaders must use ipfs-cluster-service to start the daemon that operates their IPFS Cluster node. Lastly, leaders start hosting the Gateway component, enabling authorized DApps to access VDICs via the Internet.

\subsubsection{Update VDICs}
Once created, VDICs might require updates to add and remove node operators or DApps. Such updates must be done by leaders.

\begin{figure}[!h]
    \centering
    \includegraphics[width=\linewidth]{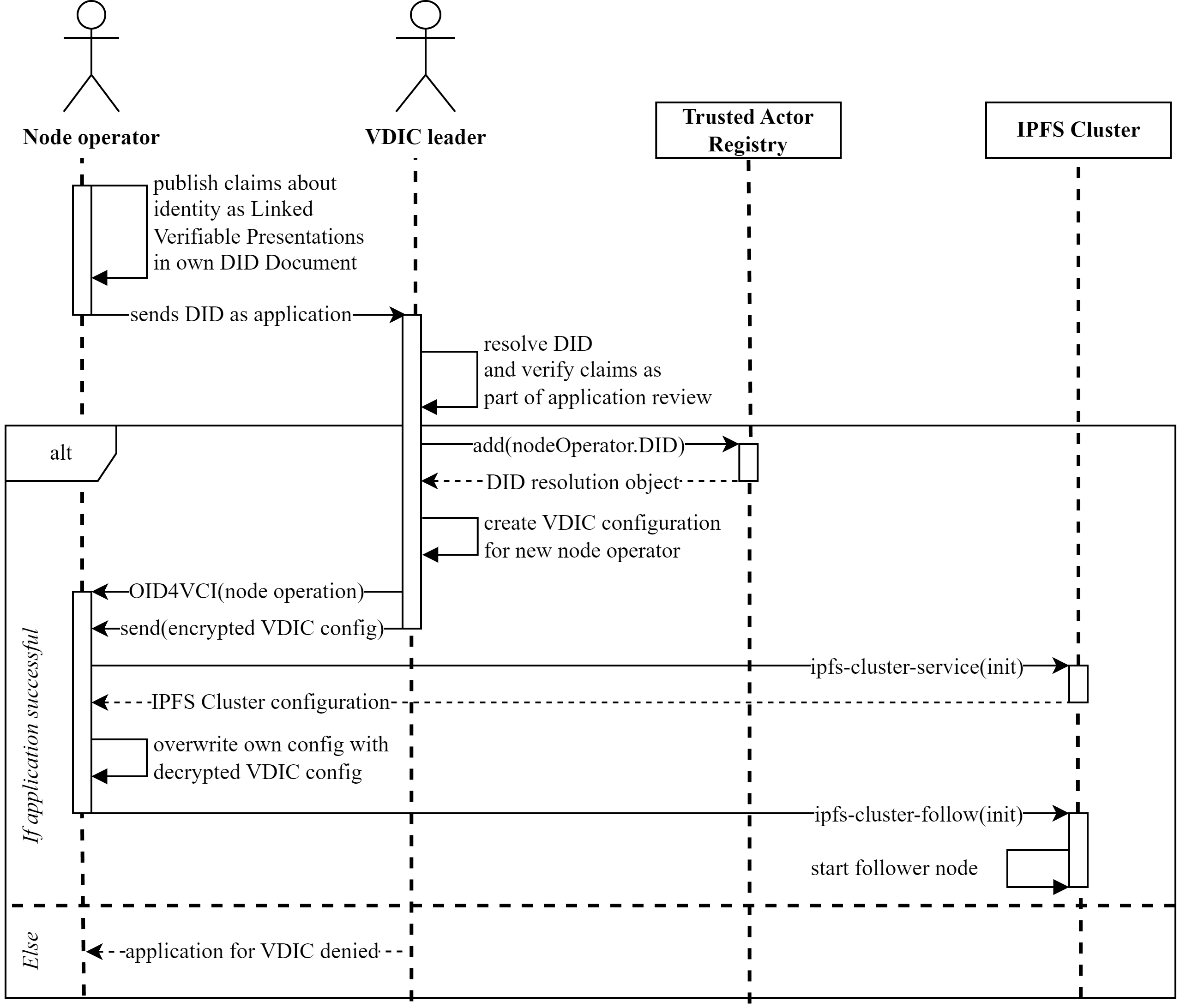}
    \caption{Adding node operators to VDICs}
    \label{fig:vdic-add-node-operator}
\end{figure}

\textit{Adding node operators} is shown in Figure~\ref{fig:vdic-add-node-operator}. Initially, leaders must perform a self-defined application process for each potential node operator. Such application processes must include requesting VCs, which prove the identity and motive for operating a node. The requested VCs could be licenses or reviews of third parties. Moreover, the application process must involve vetting the public availability of the requested VCs. Public availability is achieved by leaders mandating node operators to store the requested VCs as Linked Verifiable Presentations in their DID Document.
After successful application, leaders first update the DID Document of their VDIC to add the DID of the new node operator to the Trusted Actor Registry. Second, leaders issue VCs to authorize node operators in VDICs using existing VC protocols such as OpenID for Verifiable Credential Issuance (OID4VCI)~\cite{oid4vc}. 
Third, leaders must also send the IPFS Cluster configuration file of their VDIC to each selected node operator via secure channels. Before being sent to node operators, the IPFS Cluster configuration file is encrypted with the public key of the respective node operator, enhancing security in VDICs. The public key is attained by leaders through the resolution of the node operator's DID.
Using the decrypted IPFS Cluster configuration file, node operators can set up their existing IPFS Cluster node to follow the IPFS Cluster node of VDIC leaders, using the command line application ipfs-cluster-follow~\cite{ipfsclusterIpfsclusterfollow}. Such follower nodes only replicate data and cannot write data to VDICs. Therefore, node operators can only read the data stored in their follower node while guaranteeing data permanency for VDICs~\cite{followerNode}. 

\textit{Adding DApps} as accessors requires a successful application to VDICs by DApp providers. During application, DApp providers present their identity and motive for accessing VDICs as Linked Verifiable Presentations. After successful application, leaders issue VCs to authorize access. Lastly, leaders update the DID Document of their VDIC to add the DID of the new DApp to the Trusted Actor Registry.
    
\textit{Removing node operators} requires the revocation of their VC authorizing node operation. To revoke the VC of node operators, leaders update the Trusted Actor Registry to delist the DID of the removed node operators. Lastly, leaders need to change the secret of the IPFS Cluster underlying their VDIC, resulting in a new IPFS Cluster configuration file sent to all remaining node operators.

\textit{Removing DApps} as accessors requires the revocation of their VC for access authorization. To revoke an access VC, leaders update the Trusted Actor Registry to delist the DID of the removed DApp.

\subsubsection{Use VDICs as off-chain storage}
To read or write data in VDICs, DApps initially need to authenticate themselves to Gateways by presenting their VC for access authorization, using existing VC protocols such as OpenID for Verifiable Presentation (OID4VP)~\cite{oid4vc}. After successful authentication, the Gateway component issues one Jason Web Token (JWT) with predefined expiration time. With such a JWT, DApps can access VDICs without undergoing VC-based authentication for each request. 

\begin{figure}[!h]
    \centering
    \includegraphics[width=\linewidth]{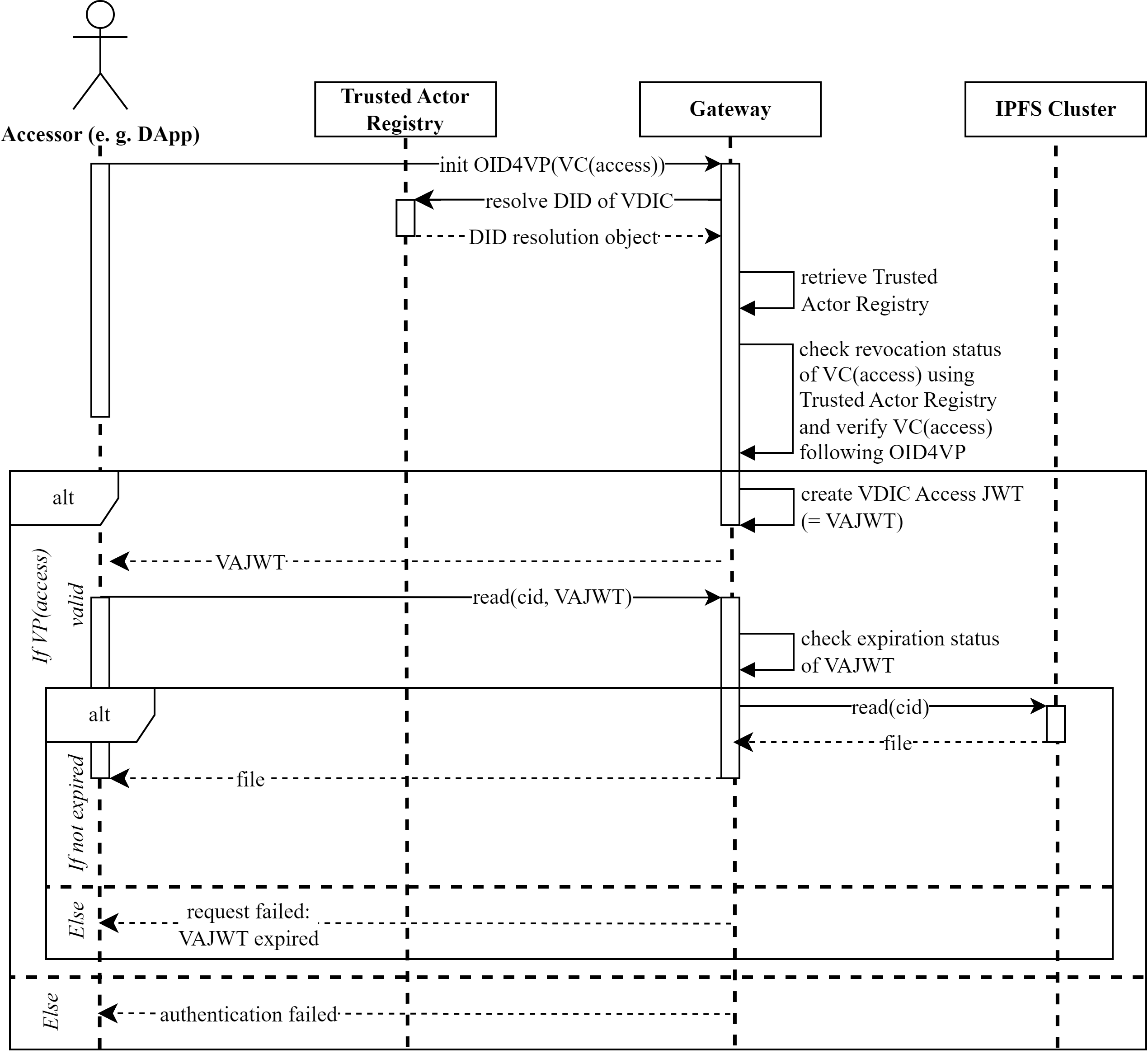}
    \caption{Reading data in VDICs} 
    \label{fig:vdic-use}
\end{figure}

Figure~\ref{fig:vdic-use} details the workflow of reading data from VDICs. The shown workflow is the same for writing data to VDICs.

\subsubsection{Configure public VDICs}
Our implemented VDIC ensures data privacy by using a private IPFS Cluster, addressing the lack of data privacy in Blockchains~\cite{8888155}. However, the common use of IPFS for off-chain storage reveals that some DApps require publicly accessible data instead of data privacy~\cite{9027313, JAYABALAN2022152, 10.1007/978-981-19-9876-8_39}. To enable public data access in VDICs, leaders can publish the IPFS Cluster configuration file, allowing anyone to operate nodes and read data. Moreover, allowing anyone to operate nodes augments the verifiability of VDICs due to enhanced auditability of other nodes' activities.

\section{Evaluation}
We evaluate VDICs to gain insights into their utility in practice by measuring their performance and exploring their use case in Supply Chain Data Management.

\subsection{Performance}
The performance of read and write operations in VDICs, compared to commonly used pinning services such as Moralis~\cite{moralis} and Pinata~\cite{pinataPinataIPFS}, is decisive for DApps considering VDICs for off-chain storage.

The GitHub repository~\cite{vdic-test-suite-repo} implements the necessary components and functionalities to test the read and write performance of VDICs. The test scripts and Gateway component are implemented in Node.js (version 20.8.1). Docker (version 25.0.3) is used to run IPFS Clusters with varying numbers of nodes. All tests are conducted on a single machine with Windows 11, an AMD Ryzen 7 7735HS processor at 3.20 GHz, and 16 GB of RAM.

\subsubsection{Read Performance}
Read performance measures the time required to retrieve data from VDICs, which is crucial for DApps needing instant data availability.

\begin{figure}[!ht]
\centering
\includegraphics[width=\linewidth]{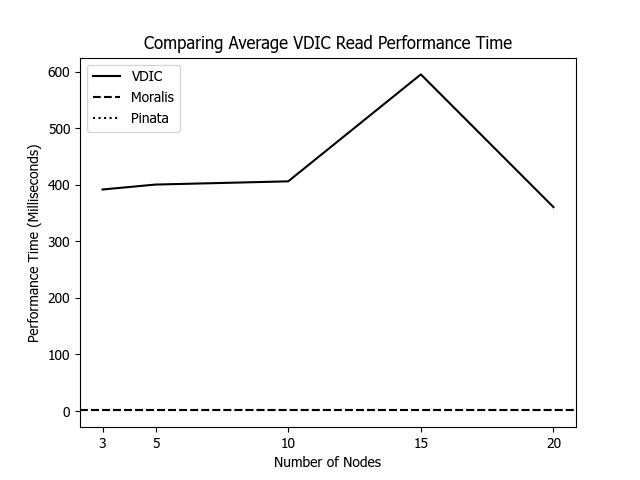}
\caption{Average read performance for a 100 KB file from VDICs with varying number of nodes compared to pinning services: Pinata and Moralis}
\label{fig-read-performance}
\end{figure}

Figure~\ref{fig-read-performance} reveals that VDICs, regardless of their number of nodes, generally exhibit slower read performance compared to the pinning services Pinata and Moralis. This slower read performance makes VDICs less favorable for DApps that demand immediate data access. Despite a slight increase in read time for VDICs with 15 nodes due to hardware and internet constraints, the average read performance for VDICs is around 400 milliseconds. The single gateway of VDICs might present a bottleneck, whereas pinning services likely have more access points and an efficient middleware to manage incoming requests. However, since pinning services do not disclose their technology or the number of nodes used for data replication, a proper comparison with VDIC is not possible. This underscores the transparency issues in pinning services.

\subsubsection{Write Performance}
Write performance measures the time needed for newly added files to replicate across all VDIC nodes.

\begin{figure}[!ht]
\centering
\includegraphics[width=\linewidth]{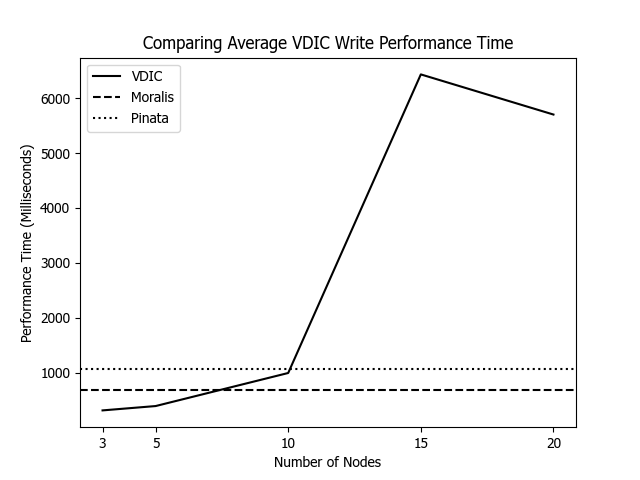}
\caption{Average write performance for a 100 KB file to VDICs with varying numbers of nodes compared to pinning services: Pinata and Moralis}
\label{fig-write-performance}
\end{figure}

Figure~\ref{fig-write-performance} shows that VDICs with up to 10 nodes achieve write performance comparable to pinning services, with average write operations taking approximately one second or less. The significant increase in write time for VDICs with 15 nodes is attributed to hardware and internet constraints. Generally, writing time increases with the number of nodes in VDICs, suggesting a trade-off between decentralization and write performance in VDICs.

\subsection{Use Case: Supply Chain Data Management}
Supply Chain Data Management (SCDM) aims to address data silos in supply chains, which arise from different entities managing their own data storage~\cite{6644169}. Dismantling data silos through unified data environments enables consumers to trace and verify product origins and compositions, which is crucial for ensuring product safety and authenticity~\cite{SUNNY2020106895}. VDICs can establish such a unified data environment for supply chains.

To use VDICs for SCDM, node operators are supply chain entities such as producers, manufacturers, and distributors. Supply chain entities are motivated by their need for data availability to enhance their productivity and data-driven processes. Additionally, regulators, governments, and NGOs can serve as node operators, focusing on compliance with laws and regulations related to fair trade, health, or sustainability. Their participation ensures real-time monitoring and accountability within the supply chain.
Leadership of VDICs in SCDM can be managed by a consortium of supply chain entities within the same domain.

\section{Discussion}
The evaluation of VDICs shows their capabilities and potential for providing performant off-chain storage with trustworthy data permanency. Despite their promising advantages, VDICs encounter challenges regarding verifiability, privacy, and security.

\subsection{Verifiability}
Validating whether node operators listed in the Trusted Actor Registry are actively operating their nodes remains challenging. The actual number of active nodes can only be verified by running a node, and leaders might overstate this number in the Trusted Actor Registry to enhance the perceived decentralization of their DApp. Requiring node operators to register their DID in the Trusted Issuer Registry of the European Blockchain Services Infrastructure (EBSI) can mitigate this issue~\cite{ebsiTrustedIssuerRegistry}. This registry lists one DID per entity, preventing leaders from overstating the number of nodes by creating multiple DIDs for non-existing node operators~\cite{ebsiIssuerTrust}. However, collusion between leaders and entities in the registry remains a concern. To address potential manipulation by leaders, VDIC operations should transition towards decentralization, distributing leadership responsibilities among actors within Decentralized Autonomous Organizations\cite{8836488}. 

\subsection{Privacy}
While VDICs offer promising privacy safeguards, some areas need improvement. One concern is the reliance on the integrity of node operators to maintain data privacy. VDICs lack mechanisms to ensure that data locally stored by node operators remains private. 

Compliance with the General Data Protection Regulation (GDPR) is also crucial for VDIC adoption in Europe. Node operators, considered data controllers, must adhere to GDPR mandates such as the "right to be forgotten"~\cite{gdprRightToBeForgotten}. Although unpinning data can facilitate deletion in VDICs, no technical restrictions prevent node operators from retaining data copies, which hinders full compliance with GDPR.

Lastly, ensuring verifiability within VDICs requires the publication of VCs held by node operators and DApps. However, when VCs are published as Linked Verifiable Presentations, holders lose control over their claims and the ability to selectively disclose them~\cite{identityLinkedVerifiable}.

\subsection{Security}
Although VDICs employ encryption to secure configuration files before transmission, a vulnerability lies in the transfer process itself, which relies on external communication channels. This dependency exposes VDICs to potential security threats. To address this risk, we propose leveraging Shared Key Generation based on Multi-party Computation. By employing Multi-party Computation protocols, VDICs can enhance their security by collaboratively generating IPFS Cluster secret keys among node operators. This method ensures that the confidentiality of the keys in VDIC configuration files remains uncompromised because keys are no longer transmitted via external communication channels.

\section{Conclusion}
Verifiable Decentralized IPFS Clusters (VDICs) advance off-chain storage solutions for Decentralized Applications. By incorporating Decentralized Identifiers and Verifiable Credentials, VDICs address the critical issues of trust and transparency in off-chain storage, providing verifiable guarantees of data permanency without relying on financial incentives or self-made proof systems. Our performance tests reveal that VDICs are competitive with existing pinning services, while the use case of Supply Chain Data Management confirms their practicality in real-world scenarios. Future work will focus on further demonstrating the practicality of VDICs by investigating their applicability in existing Decentralized Applications and exploring additional use cases, such as Decentralized Federated Learning. Additionally, efforts will be directed toward optimizing VDIC read times and advancing VDIC development based on our discussions.

\section*{Acknowledgment}
This work was supported by the European Union’s Digital Europe Program (DIGITAL) research and innovation program under grant agreement number 101102743 (TRACE4EU). The authors thank the Trace4EU project partners for their invaluable contributions.

\bibliographystyle{IEEEtran}
\bibliography{references}

\begin{thebibliography}{10}
\providecommand{\url}[1]{#1}
\csname url@samestyle\endcsname
\providecommand{\newblock}{\relax}
\providecommand{\bibinfo}[2]{#2}
\providecommand{\BIBentrySTDinterwordspacing}{\spaceskip=0pt\relax}
\providecommand{\BIBentryALTinterwordstretchfactor}{4}
\providecommand{\BIBentryALTinterwordspacing}{\spaceskip=\fontdimen2\font plus
\BIBentryALTinterwordstretchfactor\fontdimen3\font minus \fontdimen4\font\relax}
\providecommand{\BIBforeignlanguage}[2]{{%
\expandafter\ifx\csname l@#1\endcsname\relax
\typeout{** WARNING: IEEEtran.bst: No hyphenation pattern has been}%
\typeout{** loaded for the language `#1'. Using the pattern for}%
\typeout{** the default language instead.}%
\else
\language=\csname l@#1\endcsname
\fi
#2}}
\providecommand{\BIBdecl}{\relax}
\BIBdecl

\bibitem{10068327}
P.~Zheng, Z.~Jiang, J.~Wu, and Z.~Zheng, ``Blockchain-based decentralized application: A survey,'' \emph{IEEE Open Journal of the Computer Society}, vol.~4, pp. 121--133, 2023.

\bibitem{9903549}
T.~V. Doan, Y.~Psaras, J.~Ott, and V.~Bajpai, ``Toward decentralized cloud storage with ipfs: Opportunities, challenges, and future considerations,'' \emph{IEEE Internet Computing}, vol.~26, no.~6, pp. 7--15, 2022.

\bibitem{9027313}
R.~Kumar, N.~Marchang, and R.~Tripathi, ``Distributed off-chain storage of patient diagnostic reports in healthcare system using ipfs and blockchain,'' in \emph{2020 International Conference on COMmunication Systems \& NETworkS (COMSNETS)}, 2020, pp. 1--5.

\bibitem{JAYABALAN2022152}
J.~Jayabalan and N.~Jeyanthi, ``Scalable blockchain model using off-chain ipfs storage for healthcare data security and privacy,'' \emph{Journal of Parallel and Distributed Computing}, vol. 164, pp. 152--167, 2022.

\bibitem{ipfsPersistenceIPFS}
{IPFS Community}, ``{Persistence, permanence, and pinning},'' \url{https://docs.ipfs.tech/concepts/persistence}, 2024, accessed: 26-04-2024.

\bibitem{9369473}
B.~Guidi, A.~Michienzi, and L.~Ricci, ``Data persistence in decentralized social applications: The ipfs approach,'' in \emph{2021 IEEE 18th Annual Consumer Communications \& Networking Conference (CCNC)}, 2021, pp. 1--4.

\bibitem{filecoinWebsite}
{Filecoin Community}, ``{Filecoin Website},'' \url{https://filecoin.io/}, 2024, accessed: 01-06-2024.

\bibitem{Benet2017PoRep}
\BIBentryALTinterwordspacing
J.~Benet, D.~Dalrymple, and N.~Greco, ``Proof of replication,'' Protocol Labs, Tech. Rep., July 2017. [Online]. Available: \url{https://filecoin.io/proof-of-replication.pdf}
\BIBentrySTDinterwordspacing

\bibitem{proofOfSpacetime}
T.~Moran and I.~Orlov, ``Simple proofs of space-time and rational proofs of storage,'' in \emph{Advances in Cryptology - {CRYPTO} 2019 - 39th Annual International Cryptology Conference, Santa Barbara, CA, USA, August 18-22, 2019, Proceedings, Part {I}}, ser. Lecture Notes in Computer Science, A.~Boldyreva and D.~Micciancio, Eds., vol. 11692.\hskip 1em plus 0.5em minus 0.4em\relax Springer, 2019, pp. 381--409.

\bibitem{8869262}
P.~C. Bartolomeu, E.~Vieira, S.~M. Hosseini, and J.~Ferreira, ``Self-sovereign identity: Use-cases, technologies, and challenges for industrial iot,'' in \emph{2019 24th IEEE International Conference on Emerging Technologies and Factory Automation (ETFA)}, 2019, pp. 1173--1180.

\bibitem{9031543}
D.~W. Chadwick, R.~Laborde, A.~Oglaza, R.~Venant, S.~Wazan, and M.~Nijjar, ``Improved identity management with verifiable credentials and fido,'' \emph{IEEE Communications Standards Magazine}, vol.~3, no.~4, pp. 14--20, 2019.

\bibitem{8662002}
S.~Muralidharan and H.~Ko, ``An interplanetary file system (ipfs) based iot framework,'' in \emph{2019 IEEE International Conference on Consumer Electronics (ICCE)}, 2019, pp. 1--2.

\bibitem{8726493}
M.~Steichen, B.~Fiz, R.~Norvill, W.~Shbair, and R.~State, ``Blockchain-based, decentralized access control for ipfs,'' in \emph{2018 IEEE International Conference on Internet of Things (iThings) and IEEE Green Computing and Communications (GreenCom) and IEEE Cyber, Physical and Social Computing (CPSCom) and IEEE Smart Data (SmartData)}, 2018, pp. 1499--1506.

\bibitem{oid4vc}
{OpenID Foundation}, ``{OpenID for Verifiable Credentials Specification},'' \url{https://openid.net/sg/openid4vc/specifications/}, 2024, accessed: 10-06-2024.

\bibitem{openidOpenIDConnect}
------, ``{OpenID Connect},'' \url{https://openid.net/developers/how-connect-works/}, 2024, accessed: 05-06-2024.

\bibitem{ipfstech}
{IPFS Community}, ``{IPFS Website},'' \url{https://ipfs.tech/}, 2024, accessed: 21-04-2024.

\bibitem{ipfscluster}
{Protocol Labs}, ``{IPFS Cluster Website},'' \url{https://ipfscluster.io/}, 2024, accessed: 21-04-2024.

\bibitem{w3c-did-core}
\BIBentryALTinterwordspacing
M.~Sporny, A.~Guy, M.~Sabadello, and D.~Reed, ``{Decentralized Identifiers},'' W3C, Recommendation 19 July 2022, 2022. [Online]. Available: \url{https://www.w3.org/TR/2022/REC-did-core-20220719/}
\BIBentrySTDinterwordspacing

\bibitem{w3c-vc-data-model}
\BIBentryALTinterwordspacing
{M. Sporny, D. Longley, D. Chadwick, and O. Steele}, ``{Verifiable Credentials Data Model v2.0},'' W3C, Candidate Recommendation Draft 16 April 2024, 2024. [Online]. Available: \url{https://www.w3.org/TR/2024/CRD-vc-data-model-2.0-20240416/}
\BIBentrySTDinterwordspacing

\bibitem{identityLinkedVerifiable}
B.~R. Jan Christoph~Ebersbach and M.~Sabadello, ``{Linked Verifiable Presentation v0.4.0},'' \url{https://identity.foundation/linked-vp/?ref=blog.identity.foundation#linked-verifiable-presentation}, 2024, accessed: 21-05-2024.

\bibitem{ipfsclusterREST}
{Protocol Labs}, ``{IPFS Cluster Documentation},'' \url{https://ipfscluster.io/documentation/reference/api/}, 2024, accessed: 23-05-2024.

\bibitem{ipfsclusterIpfsclusterservice}
------, ``{ipfs-cluster-service Documentation},'' \url{https://ipfscluster.io/documentation/reference/service/}, 2024, accessed: 27-05-2024.

\bibitem{ipfsclusterIpfsclusterfollow}
------, ``{ipfs-cluster-follow Documentation},'' \url{https://ipfscluster.io/documentation/reference/follow/}, 2024, accessed: 27-05-2024.

\bibitem{followerNode}
------, ``{Joining a collaborative cluster},'' \url{https://ipfscluster.io/documentation/collaborative/joining/}, 2024, accessed: 27-05-2024.

\bibitem{8888155}
J.~Bernal~Bernabe, J.~L. Canovas, J.~L. Hernandez-Ramos, R.~Torres~Moreno, and A.~Skarmeta, ``Privacy-preserving solutions for blockchain: Review and challenges,'' \emph{IEEE Access}, vol.~7, pp. 164\,908--164\,940, 2019.

\bibitem{10.1007/978-981-19-9876-8_39}
M.~Kaur, S.~Gupta, D.~Kumar, M.~S. Raboaca, S.~B. Goyal, and C.~Verma, ``Ipfs: An off-chain storage solution for blockchain,'' in \emph{Proceedings of International Conference on Recent Innovations in Computing}, Y.~Singh, P.~K. Singh, M.~H. Kolekar, A.~K. Kar, and P.~J.~S. Gon{\c{c}}alves, Eds.\hskip 1em plus 0.5em minus 0.4em\relax Singapore: Springer Nature Singapore, 2023, pp. 513--525.

\bibitem{moralis}
{Moralis Web3 Technology AB}, ``{How to Upload Files to IPFS},'' \url{https://moralis.io/how-to-upload-files-to-ipfs-full-guide/}, 2022, accessed: 24-07-2024.

\bibitem{pinataPinataIPFS}
{Pinata Technologies, Inc.}, ``{Pinata Website},'' \url{https://www.pinata.cloud/}, 2024, accessed: 30-05-2024.

\bibitem{vdic-test-suite-repo}
S.~Lamichhane, ``{Evaluation VDIC},'' \url{https://github.com/sid030sid/evaluation-vdic}, 2024, accessed: 19-05-2024.

\bibitem{6644169}
S.~Robak, B.~Franczyk, and M.~Robak, ``Applying big data and linked data concepts in supply chains management,'' in \emph{2013 Federated Conference on Computer Science and Information Systems}, 2013, pp. 1215--1221.

\bibitem{SUNNY2020106895}
J.~Sunny, N.~Undralla, and V.~M. Pillai, ``{Supply chain transparency through blockchain-based traceability: An overview with demonstration},'' \emph{Comput. Ind. Eng.}, vol. 150, p. 106895, 2020.

\bibitem{ebsiTrustedIssuerRegistry}
{European Blockchain Services Infrastructure}, ``{Issuer Issuer Registry},'' \url{https://hub.ebsi.eu/apis/conformance/trusted-issuers-registry/v4}, 2024, accessed: 05-06-2024.

\bibitem{ebsiIssuerTrust}
------, ``{Issuer Trust model},'' \url{https://hub.ebsi.eu/vc-framework/trust-model/issuer-trust-model-v4}, 2024, accessed: 05-06-2024.

\bibitem{8836488}
S.~Wang, W.~Ding, J.~Li, Y.~Yuan, L.~Ouyang, and F.-Y. Wang, ``Decentralized autonomous organizations: Concept, model, and applications,'' \emph{IEEE Transactions on Computational Social Systems}, vol.~6, no.~5, pp. 870--878, 2019.

\bibitem{gdprRightToBeForgotten}
{Proton AG}, ``{Right to be forgotten},'' \url{https://gdpr.eu/right-to-be-forgotten/}, 2024, accessed: 10-06-2024.

\end{thebibliography}

\end{document}